\documentclass[twoside,12pt]{article}
\usepackage{latexsym,amsmath}
\usepackage{amssymb}
\usepackage{hyperref}
\usepackage{amsfonts}
\usepackage{color}
\usepackage{amsmath}
\usepackage{dsfont}
\usepackage{enumerate}
\usepackage{natbib}
\usepackage{graphics,graphicx}
\usepackage[center]{titlesec}
\begin{document}
\setcounter{page}{1}
\begin{center}

\vspace{0.4cm} {\large{\bf\bf Kaluza--Klein String Cosmological Model in $f(R,T)$ Theory of Gravity
}} \\

\vspace{0.4cm}
D. D. Pawar, G. G. Bhuttampalle and P. K. Agrawal \\
School of Mathematical Sciences,\\
Swami Ramanand Teerth Marathwada University, Nanded-431606, India\\
E-mail: dypawar@yahoo.com, bgovardhan14@gmail.com and agrawalpoonam299@gmail.com
\\
\vspace {0.2cm}
\end{center}
\begin{abstract}
 In this paper we have studied Kaluza--Klein string cosmological model within the framework of $f(R,T)$ theory of gravity, where $R$ is the Ricci scalar and $T$ is the trace of the stress energy momentum tensor. We have obtained the solution of the corresponding field equations by using a time varying deceleration parameter. We also discussed various physical and dynamical properties of the model. The variation of different cosmological parameters are shown graphically for specific values of the parameters of the model.
\end{abstract}
\def\baselinestretch{1.5}
\allowdisplaybreaks
%\begin{center}
\section{Introduction}
%\end{center}

Einstein's general theory of relativity (GR) explains large number of gravitational phenomena such as bending of light, motion of planet Mercury, expansion of the Universe, etc. GR also predicts the presence of gravitational waves which have been recently detected by international collaboration LIGO. 
%The determined results of GR are obeyed %(approved) with the standard model of particles. 
The behavior of solar system like the elliptical orbits of planets/comets/moons around the Sun, their periodicity was possible to explain using the simplest Newtonian mechanics but the complex behavior was possible only after the advancement in the theory of general relativity. Despite these success of GR there are few drawbacks e.g. it fails to explain the accelerated expansion of the Universe. The late time acceleration is usually associated with a strong negative pressure in the form of exotic (and mostly unknown) dark energy (DE). There are many candidates of DE such as the cosmological constant (may be time varying), quintessence, phantom, quantum, etc. which are not explained anywhere in the GR \citep{doi:10.1063/1.1724264,2009FoPh...39.1161C}. At present, it is considered that the Universe is mostly dictated by the presence of 68.3\% dark energy, 26.8\% dark matter and 4.9\% of baryonic matter.
In order to study the DE and the consequent cosmic acceleration, several modified theories of gravity have been evolved like, $f(R)$ gravity, $f(R,T)$ gravity, $f(T)$, $f(G)$, $f(R,G)$ gravity, etc. Among these theories, in our study, we considered $f(R,T)$ gravity.\\

$f(R,T)$ theory of gravity is developed by \cite{PhysRevD.84.024020}, where the gravitational Lagrangian is a function of $R$ and trace $T$ of the energy momentum tensor. The dependence on $T$ may be induced by exotic imperfect fluid and/or certain quantum effects. These models also depend on the variation of the matter stress energy tensor \citep{Ram2014a}. $f(R,T)$ theory made advancement in GR due to the coupling of the matter and geometry. Here, the covariant divergence of the stress energy tensor is non--zero. As a result, the motion of test particles is not along geodesic path. The late time cosmic accelerated expansion of the universe can be explain by $f(R,T)$ gravity models. The field equations of $f(R,T)$ gravity can be determined by varying the action of the gravitational field equations with respect to the metric tensor. Several authors e.g. \cite{2012EPJC...72.2203M,Reddy2012,Chaubey2013,Ram2014,Pawar2015a,Chirde2015,Pawar2016,Pawar2016b,2017NewA...54...56A}, etc. have studied various cosmological models in the framework of $f(R,T)$ gravity theory. \cite{Samanta2013} obtained the exact solutions of Kantowski--Sachs cosmological model filled with perfect fluid matter in the presence of $f(R,T)$ gravity. \cite{Mishra2014} derived Bianchi type $VI_{h}$ cosmological model filled with perfect fluid in $f(R,T)$ gravity. \cite{2017arXiv170706968M} have discussed the modeling of static wormholes in the framework of $f(R, T)$ gravity.\\

$f(R,T)$ gravity is also useful in string cosmology models.  The study of string cosmological models (as cosmic strings) find considerable attention in cosmology. It is assumed that cosmic strings play pivotal role in the early evolution of the universe, particularly before the particle creation. Grand unified theories anticipate such strings to be formed during the phase transition after the big--bang explosion at the temperature less than the critical temperature. Furthermore, it is predicted that cosmic strings are linear topological defects associated with spontaneous symmetry breaking whose plausible production site is cosmological phase transitions in the early universe. They also considered to be acting as a gravitational lens and therefore are assumed as possible seeds for the formation of galaxies. \cite{PhysRevD.20.1294,PhysRevD.21.2171,PhysRevD.24.2082,PhysRevD.28.2414,1985ApJ...288..422G} have broadly discussed the gravitational effects of cosmic strings in general relativity. \cite{Krori1990,Tikekar1992} have examined Bianchi type space--time for a cloud string. \cite{2007ChPhL..24..585R} have determined Bianchi type--$III$ string cosmological models in the presence of bulk viscous fluid for massive string. \cite{Pawar2008} discussed dust magnetized string cosmological model while \cite{Rao2008} investigated the string cosmological model in Saez--Ballester theory of gravitation. \cite{Sahoo2016} have discussed Bianchi--$III$ and --$VI_{0}$ string fluid source cosmological models in the framework of $f(R,T)$ gravity. Beside this, \cite{pawar2010} have elaborated plane symmetric string cosmological model with bulk viscosity and \cite{pawar2012} have discussed the plane symmetric cosmological models with string dust magnetized bulk viscosity in Lyra geometry. \cite{Sahoo2013} studied the plane symmetric cosmological solutions for quark matter with the string cloud and domain walls using Rosen’s bimetric theory. \cite{PhysRevD.21.2167,PhysRevD.30.344,Chatterjee1993} have studied higher dimensional string cosmological models in the context of different theories.\\

Generally, with the usual (four dimensional) space--time, the consolidation of gravitational forces with other forces of nature is not possible. Thus to improve the possibility of geometrically unifying the fundamental interactions of the universe, study of higher dimensional space--time is essential. In particle physics, various experiments were initiated to develop the higher dimensional cosmological models that resulted into the formulation of the Kaluza--Klein theory. Kaluza--Klein (KK) theories reveal how the gravity and the electromagnetism can be unified from Einstein's field equations generalized to five dimensions. Subsequently, different authors studied physics of the universe in the context of higher dimensional space--time viz. \cite{PhysRevLett.51.931,RANDJBARDAEMI1984388,PhysRevLett.52.489} have shown that the experimental detection of time variation of fundamental constants could provide evidence for extra space dimensions. Moreover, multi--dimensional cosmological models are studied by several authors in the framework of different theories \cite{PhysRevD.21.2167,Lorenz-Petzold1985,PhysRevD.34.1202,PhysRevD.37.3761,Reddy2001}. \cite{Namrata2013} have obtained exact solutions of Einstein field equations of Kaluza--Klein cosmological model with strange quark matter and string cloud. \cite{SSPawar2015} have studied Kaluza--Klein cosmological model in the presence of $f(R,T)$ theory of gravity. \cite{KATORE2015172} have determined Kaluza--Klein universe in the presence of magnetized dark energy in the reference of Lyra manifold.\\

Motivated by the above works in the cosmology, here we present the study of Kaluza--Klein string cosmological model in the $f(R,T)$ theory of gravity. In section 4, we obtained the solutions of the field equations by considering a power--law relation between the scale factors and the special law of variation of Hubble's parameter proposed by \cite{Berman}. In the subsequent section, we derived few physical parameters of the model. In the last section, we discuss the physical behavior of the model with the help of the derived physical parameters.

\section{Gravitational field equations of f(R,T) gravity}
\cite{PhysRevD.84.024020} has obtained the field equations of $f(R,T)$ theory of gravity from the Hilbert-Einstein variational principle by considering the metric-dependent Lagrangian density $L_{m}$. The action for $f(R,T)$ gravity is given as,
\begin{equation}\label{eq:x1}
S = \int \frac{1}{16 \pi}f(R,T)\sqrt{-g}d^4x + \int L_m\sqrt{-g}d^4x,
\end{equation}
where $L_m$ is the matter Lagrangian density and $f(R,T)$ is function of Ricci scalar $R$ and $T$. $T$ is the trace of energy momentum tensor of the matter $T_{ij}$ which is given by,
\begin{equation}\label{eq:x2}
T_{ij}= \frac{-2\partial(\sqrt{-g}L_m)}{\sqrt{-g}\partial g^{ij}},
\end{equation}
and the trace of energy momentum tensor is $ T=g^{ij}T_{ij}$.
Here, it is considered that the matter Lagrangian $ L_m$ depends on the metric tensor component $ g_{ij} $ rather than its derivatives. Which implies,  
\begin{equation}\label{eq:x3}
T_{ij} = g_{ij}L_{m} - \frac{\partial L_{m}}{\partial g^{ij}}.
\end{equation}
The field equations of $f(R,T)$ gravity are given by varying the action $S$ with respect to metric tensor $g_{ij}$
\begin{equation}\label{eq:x4}
f_R(R,T)R_{ij} -\frac{1}{2}f(R,T)g_{ij} + (g_{ij}\square - \bigtriangledown_{i}\bigtriangledown_{j})f_R(R,T) =  8\pi T_{ij} - f_T(R,T)T_{ij} - f_T(R,T)\Theta _{ij},
\end{equation}
where the value of $\Theta _{ij}$ is,
\begin{equation}\label{eq:x5}
\Theta_{ij} = -2T_{ij} + g_{ij}L_{m} -2g^{l\alpha } \frac{\partial^{2}L_{m}}{\partial g^{ij}\partial g^{l\alpha }}.
\end{equation}
Here $f_R(R,T)= \frac{\partial f(R,T)}{\partial R}, f_T(R,T) = \frac{\partial f(R,T)}{\partial T}, \square =\triangledown ^{n}\triangledown_{n}$ where $\triangledown_{n}$ is the covariant derivative.\\
By contracting eqn~\eqref{eq:x4}, we obtain
\begin{equation}\label{eq:x6}
f_R(R,T)R + 3\square f_R(R,T) - 2f(R,T)= 8 \pi T - f_T(R,T)(T+ \theta),
\end{equation}
where $\Theta = g^{ij}\Theta_{ij}$.\\
The stress energy tensor of the matter is obtained by using matter Lagrangian $L_m$ as,
\begin{equation}\label{eq:x7}
T_{ij} = (\rho + p)u_{i}u_{j} +pg_{ij}-\lambda x_ix_j,
\end{equation}
where $\rho$ is the energy density of the fluid and $p$ is the pressure of the fluid. Here, $ u^{i} = (0,0,0,0,1)$ is the five-velocity vector in co--moving co--ordinate system and satisfies the condition $u_{i}u_{j} = -x_{i}x_{j}=-1$, $u_{i}x_{i}=0$ where $x^i=(0,0,0,\frac{1}{\beta},0)$ is the direction of the string and $ u^{i}\triangledown_{j} u_{i} = 0$. We choose a perfect fluid matter characterized by $L_{m} = -p$ to obtain
\begin{equation}\label{eq:x8}
\Theta_{ij} = -2T_{ij} -pg_{ij}.
\end{equation}
The field equations of $f(R,T)$ gravity depends on the tensor $\Theta_{ij}$. Thus depending on the nature of the matter source, various cosmological models of the $f(R,T)$ gravity are possible. Initially, \cite{PhysRevD.84.024020} derived three classes of models using following functional forms of $f$,
\begin{equation}\label{eq:x9}
%\begin{cases} 3x + 5y + z \\ 7x – 2y + 4z \\ -6x + 3y + 2z \end{cases}
f(R,T)= \begin{cases}  R + 2f(T) \\ f_{1}(R) +f_{2}(T) \\ f_{1}(R) +f_{2}(R)f_{3}(T).\end{cases}
\end{equation} 
{\bf Harko argued that, the field equations depend on the physical nature of the matter field. Hence depending on the nature of the matter source, for each choice of $f$ we can obtain several theoretical models, corresponding to different matter models.} Among these, we considered second class, i. e. $f(R,T) =f_{1}(R) + f_{2}(T)$ with $f(T)=\mu T $ and $f(R)=\mu R$, where $f(T)$ is the arbitrary function of stress energy of matter, $f(R)$ is the arbitrary function of the Ricci scalar $R$ and $\mu$ is an arbitrary constant. {\bf We choose this simple form as it becomes $f(R)$ if $f_{2}(T)=0$ and also it makes calculations more simpler.} Eqn~\eqref{eq:x4} gives the gravitational field equations of $f(R,T)$ gravity as follows,
\begin{equation}\label{eq:x10}
R_{ij} - \frac{1}{2}Rg_{ij} =\left(\frac{ 8 \pi +\mu}{\mu}\right)T_{ij} +\left(p+\frac{1}{2}T\right)g_{ij}.
\end{equation}
\section{Metric and the field equations}
We consider the Kaluza--Klein metric as
\begin{equation}\label{eq:x11}
ds^{2} = dt^{2} - A^{2}(dx^{2} + dy^{2}+ dz^{2})- B^{2}du^{2},
\end{equation}
where the cosmic scale factors $A$ and $B$ are the functions of $t$ and the fifth co--ordinate $u$ is taken to be space--like. The spatial volume $V$ of the universe is defined as,
\begin{equation}\label{eq:x12}
V = a^{4}= A^{3}B,
\end{equation}
where $a$ is the mean scale factor. The generalized mean Hubble's parameter $H$ for Kaluza--Klein space--time is given by,
\begin{equation}\label{eq:x13}
H = \frac{1}{4} \left(\frac{3\dot A}{A}+\frac{\dot B}{B}\right),
\end{equation}
where overhead dot denotes derivatives with respect to time $t$. The directional Hubble parameters in the direction of $x$, $y$, $z$ and $u$ are
\begin{equation}\
H_{x}=H_{y}=H_{z} = \frac{\dot{A}}{A} \quad \rm{and} \quad H_{u}=\frac{\dot{B}}{B}.
\end{equation}
The expansion scalar $\theta$ is given as,  
\begin{equation}\label{eq:x14}
\theta = 4H = \frac{3\dot A }{A}+ \frac{\dot B}{B}.
\end{equation}
The shear scalar $\sigma$ and the mean anisotropy parameter $\Delta$ are defined as,
\begin{equation}\label{eq:x15}
\sigma^{2} = \frac{1}{2}\left[\sum_{i=1}^{4} H_{i}^{2} -4 H^{2}\right]= \frac{4}{2}\Delta H^{2},
\end{equation}
and
\begin{equation}\label{eq:x16}
\Delta = \frac{1}{4}\sum_{i=1}^{4} \left(\frac{\triangle H{i}}{H}\right)^{2},
\end{equation}
where $\triangle H_{i} = H_{i} - H.$\\
The field eqn~\eqref{eq:x10} with energy momentum tensor eqn~\eqref{eq:x7} for the metric eqn~\eqref{eq:x11} takes the form,
\begin{equation}\label{eq:x17}
\left({\frac{\dot{A}}{A}}\right)^{2} +2 \left(\frac{\dot{A}\dot{B}}{AB}\right)+2\left(\frac{\ddot{A}}{A}\right) = p(8\pi+10\mu) +\lambda\mu +\mu\rho,
\end{equation}
\begin{equation}
\left({\frac{\dot{A}}{A}}\right)^{2} +2 \left(\frac{\dot{A}\dot{B}}{AB}\right)+2\left(\frac{\ddot{A}}{A}\right)= p(8\pi+10\mu) +\lambda\mu +\mu\rho,
\end{equation}

\begin{equation}\label{eq:x18}
3\left(\frac{\ddot{A}}{A}\right)^2 +3\left(\frac{\ddot{A}}{A}\right) = p(8\pi+10\mu)  +\mu\rho +\lambda(8\pi+3\mu),
\end{equation}
\begin{equation}\label{eq:x19}
-3\left(\frac{\dot{A}}{A}\right)^2 +3\left(\frac{\dot{A}\dot{B}}{AB}\right) = p(16\pi+12\mu) +\rho [8\pi+3\mu]+\lambda\mu.
\end{equation}
\section{Solution of the field equations}
    Now we have a set of three nonlinear equations with five unknowns $A$, $B$, $p$, $\rho$ and $\lambda$. Therefore to find a consistent solution from these equations we need atleast two assumptions to simplify the mathematics. Consider the special law of variation of Hubble's parameter proposed by \cite{Berman} which yields the constant deceleration parameter given by the relation,
\begin{equation}\label{eq:x20}
q = -\frac{a\ddot a}{\dot a^2} .
\end{equation}
In eqn~\eqref{eq:x20}, $a$ is an average scale factor. For a given metric (eqn~\eqref{eq:x11}), $a$ is given by,
\begin{equation}\label{eq:x21}
V =a^4 \quad \Rightarrow \quad a=V^\frac{1}{4}=(A^3B)^\frac{1}{4}.
\end{equation}
Solving the eqn~\eqref{eq:x20} we get the deceleration parameter $q$ as,
\begin{equation}\label{eq:x22}
q = \frac{-\alpha}{t^2} +(\beta-1).
\end{equation}
This equation produces a constant value for deceleration parameter and can have both positive and negative values. The positive values of deceleration parameter gives the standard deceleration model while the negative value results into inflation or the accelerating universe. On integrating eqn~\eqref{eq:x20} we get a scale factor $a$ as,
\begin{equation}\label{eq:x23}
a(t)= e^{\lambda} exp{\int \frac{dt}{\int(1+q)dt+\delta}},
\end{equation}
where $\delta$, $\lambda$ are integrating constants, eqn~\eqref{eq:x23} can not be integrated for different values of constants. Solving the eqn~\eqref{eq:x23} with the help of eqn~\eqref{eq:x22} and considering $\delta=\lambda=0$ we get,
\begin{equation}\label{eq:x24}
a(t) = \left(t^2+\frac{\alpha}{\beta} \right)^\frac{1}{2\beta}.
\end{equation}       
To solve the system completely, we assume shear scalar $\sigma$ to be proportional to the expansion scalar $\theta$. This gives a linear relationship between the Hubble's parameters $H_x$ and $H_z$. This assumption gives an anisotropic relation between the scale factors $A$ and $B$ as, $ A = B^n $
where $n \neq 1$ is the arbitrary constant. Now, using $\alpha =0$ in eqn~\eqref{eq:x24} and by eqn~\eqref{eq:x21} we get, 
\begin{equation}
t^\frac{4}{\beta} = B^{3n+1},
\end{equation}
\begin{equation}\label{eq:x25}
B = t^\frac{4}{\beta(3n+1)} \quad \rm{and} \quad A = t^\frac{4n}{\beta(3n+1)} .
\end{equation}
%\begin{equation}\label{eq:x26}
%\end{equation}
From eqn~\eqref{eq:x17} and eqn~\eqref{eq:x18} we get the expression for $\lambda$ as,
\begin{equation}\label{eq:x27}
\lambda=\frac{6\beta n^2+8\beta n-16n+8n^2-8+2\beta}{(4\pi+\mu)\beta^2(3n+1)^2t^2}.
\end{equation}

\section{Some Physical Parameters of the Model}
    In this section, we discuss the physical properties of the model by deriving solutions of few physical parameters. The spatial volume $V$ of the universe is given as,
\begin{equation}\label{eq:x28}
V=a^4=A^3B.
\end{equation}
The mean generalized Hubble’s parameter $H$ is defined as,
\begin{equation}\label{eq:x29}
H=\frac{1}{4}\left(\frac{3\dot{A}}{A}+\frac{\dot{B}}{B}\right),
\end{equation}
and for the present model it is found to be equal to,
\begin{equation}\label{eq:x30}
H=\frac{1}{4t}\left[\frac{3n+1}{\beta n}\right].
\end{equation}
The expansion scalar $\theta$ of the model is given by,
\begin{equation}\label{eq:x31}
\theta=\frac{3}{4}\left(\frac{3\dot{A}}{A}+\frac{\dot{B}}{B}\right),
\end{equation}
and here it becomes
\begin{equation}\label{eq:x32}
\theta=\frac{3}{4t}\left[\frac{3n+1}{\beta n}\right].
\end{equation}
Shear scalar $\sigma$ of the model is defined as,
\begin{equation}\label{eq:x33}
\sigma^2=\frac{3}{4}\theta^2,
\end{equation}
and found as
\begin{equation}\label{eq:x34}
\sigma^2=\frac{27}{64t^2}\frac{(3n+1)^2}{\beta^2n^2}.
\end{equation}
Deceleration parameter is given by,
\begin{equation}\label{eq:x35}
q=\frac{d}{dt}\left(\frac{1}{H}-1\right),
\end{equation}
and evaluated as
\begin{equation}\label{eq:x36}
q=\frac{4\beta n}{(3n+1)}-1.
\end{equation}
The anisotropic parameter is given by,
\begin{equation}
\Delta = \frac{3(n-1)^2}{(3n+1)^2}.
\end{equation}
We also tried to obtain the pressure and energy density of the model. After simple mathematical calculations we obtained them to be,
\begin{align}\label{eq:x37}
p &= \frac{-384\pi^2\beta n^2-512\pi^2\beta n+1024\pi^2n+1024\pi^2n^2+512\pi^2-128\pi^2\beta-432\pi\mu\beta n^2} {(4\pi+\mu)+\beta^2(3n+1)^2t^2(64\pi^2+88\pi\mu+18\mu^2)}   \nonumber \\
& - \frac{432\pi\mu\beta n+576\pi\mu n-64\pi\mu n^2+384\pi\mu-84\mu^2\beta n^2-76\mu^2\beta n+80\mu^2n+80\mu^2 n^2} {(4\pi+\mu)+\beta^2(3n+1)^2t^2(64\pi^2+88\pi\mu+18\mu^2)} \nonumber \\
&+\frac{64\mu^2-16\mu^2\beta}{(4\pi+\mu)+\beta^2(3n+1)^2t^2(64\pi^2+88\pi\mu+18\mu^2)},
\end{align}
\begin{align}\label{eq:x38}
	\rho & = \left[\frac{-12\beta n^2-16\beta n+32n+32n^2+16-4\beta}{\beta^2(3n_1)^2t^2}\right]-\left[\frac{6\beta n^2+8\beta n-16n+8n^2-8+2\beta}{(4\pi+\mu)\beta^2(3n+1)^2t^2}\right] \nonumber\\
& -(8\pi+10\mu)\left[\frac{-384\pi^2\beta n^2-512\pi^2\beta n+1024\pi^2n+1024\pi^2n^2+512\pi^2-128\pi^2\beta - 432\pi\mu\beta n^2} {(4\pi+\mu)+\beta^2(3n+1)^2t^2(64\pi^2+88\pi\mu+18\mu^2)}\right] \nonumber\\
& + (8\pi+10\mu)\left[\frac{432\pi\mu\beta n+576\pi\mu n-64\pi\mu n^2+384\pi\mu-84\mu^2\beta n^2-76\mu^2\beta n+80\mu^2n} {(4\pi+\mu)+\beta^2(3n+1)^2t^2(64\pi^2+88\pi\mu+18\mu^2)}\right] \nonumber\\
&-(8\pi+10\mu)\left[\frac{	80\mu^2 n^2+64\mu^2-16\mu^2\beta}{(4\pi+\mu)+\beta^2(3n+1)^2t^2(64\pi^2+88\pi\mu+18\mu^2)}\right].
\end{align}

\begin{figure}
\includegraphics[scale=0.45]{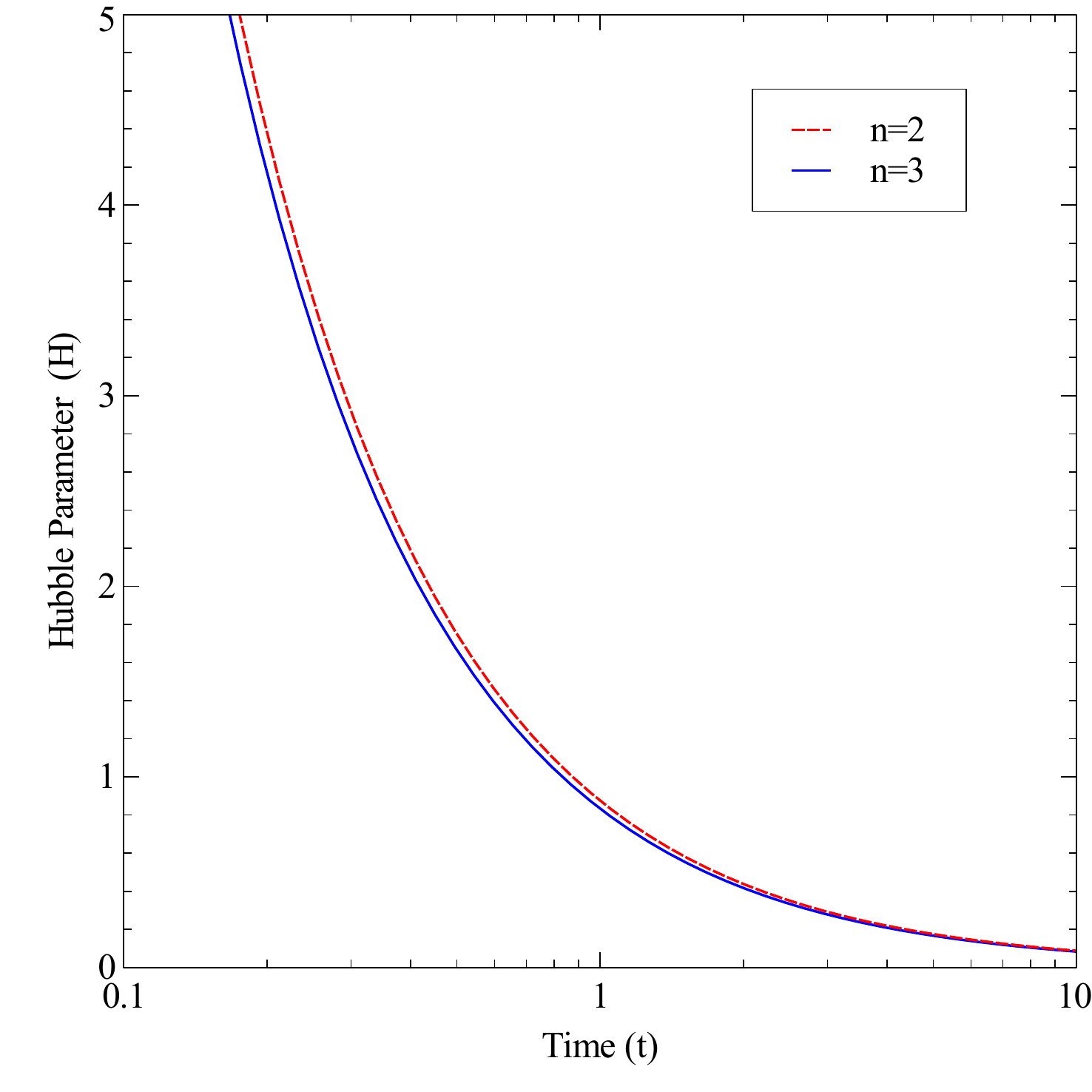}
\includegraphics[scale=0.45]{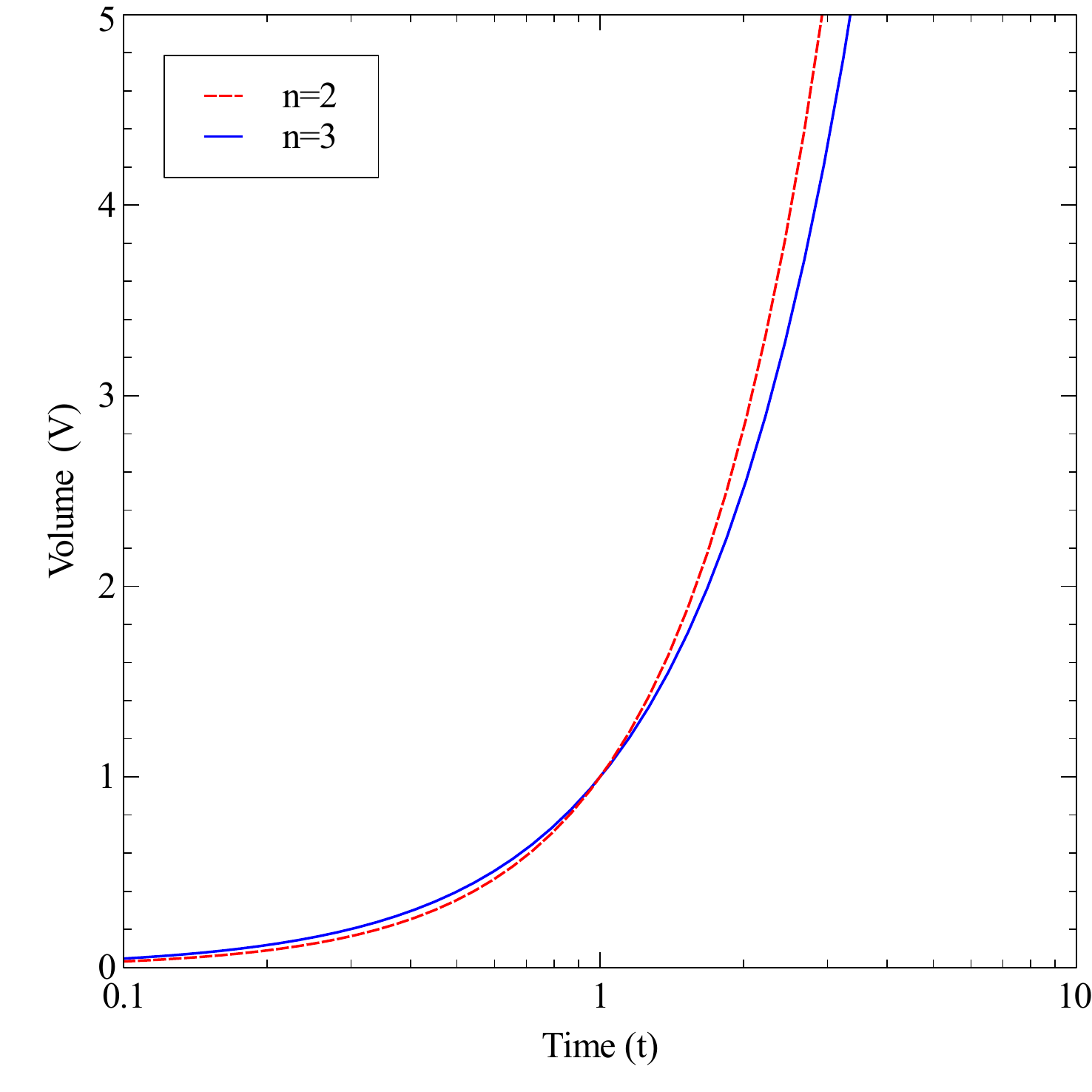}
\caption{Variations of Hubble's parameter $H$ (\emph{left}) and of volume $V$ (\emph{right}) as a function of time $t$. All quantities are in arbitrary units.}\label{fig1}
\end{figure}
\begin{figure}
%\vspace{1cm}
\includegraphics[scale=0.45]{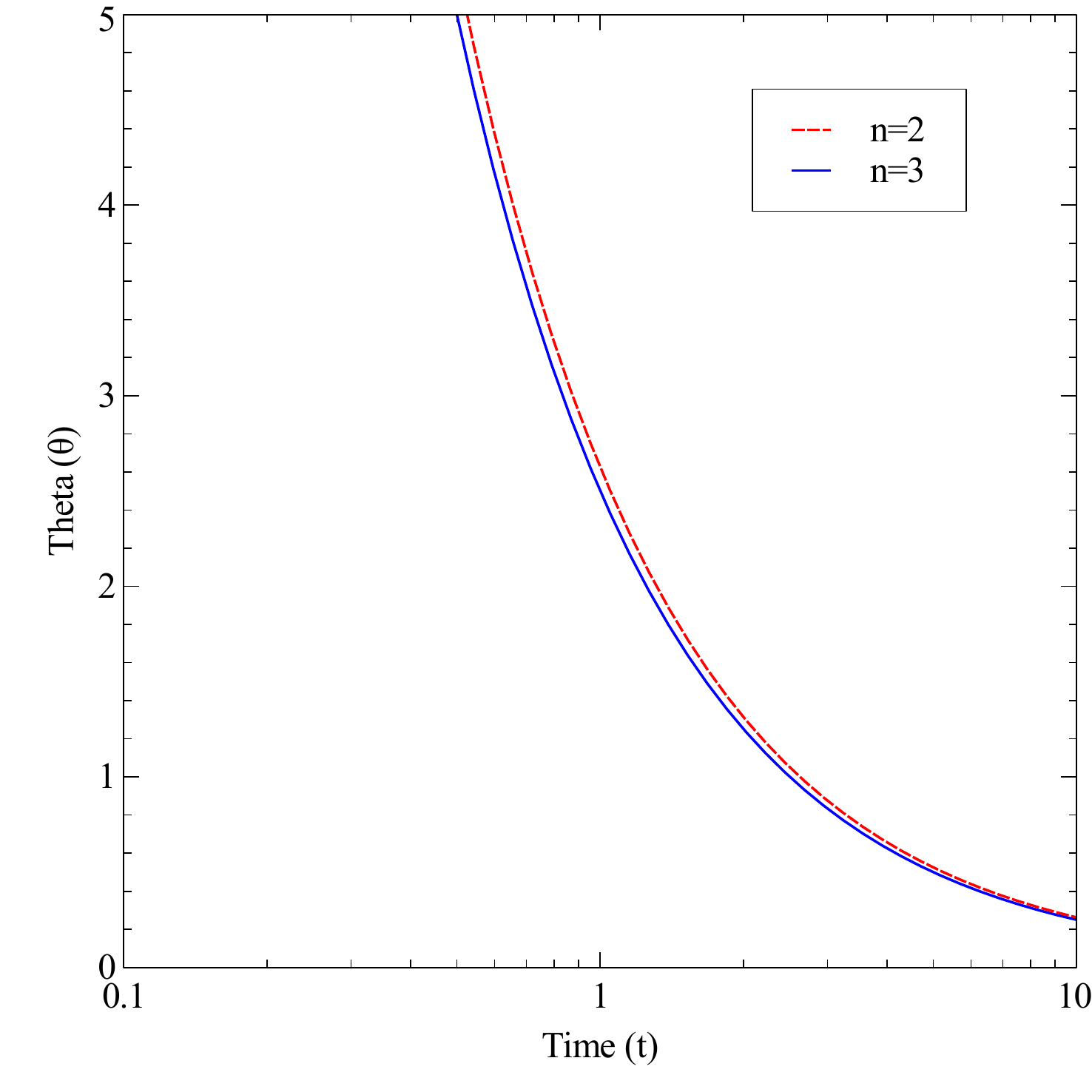}
\includegraphics[scale=0.45]{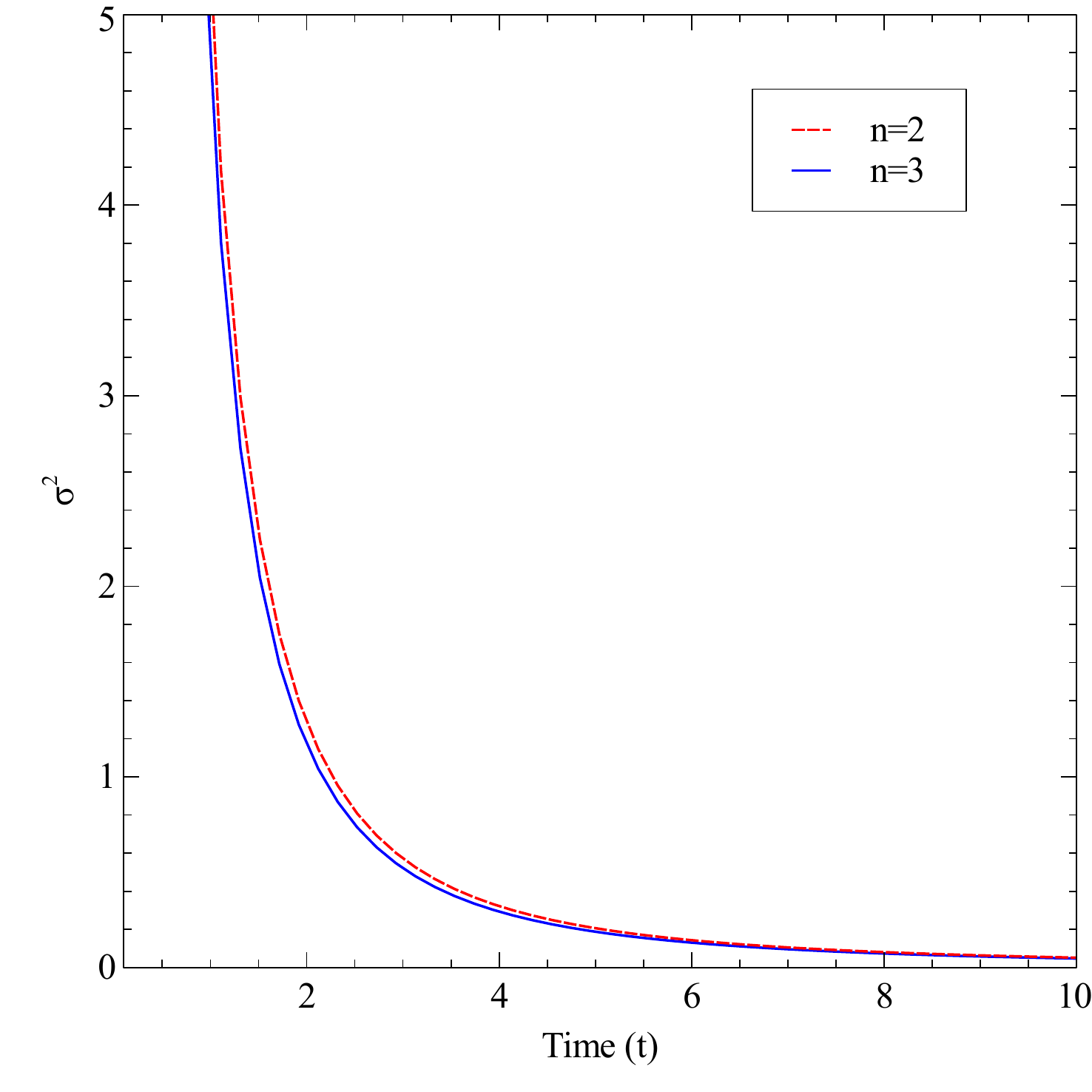}
\caption{Variations of scalar expansion $\theta$ (\emph{left}) and sigma $\sigma$ (\emph{right}) as a function of time $t$. $\theta$, $\sigma$ and $t$ are in arbitrary units.}\label{fig2}
\end{figure}

\section{Discussion and conclusion}

In this paper, we studied the Kaluza--Klein string cosmological model in the framework of $f(R,T)$ theory of gravity. To find the corresponding field equations, we assumed a power law relation between the scale factors. We also assumed a time varying deceleration parameter. The physical behavior of the solutions derived in the previous section are discussed here. The equations ~\eqref{eq:x24}, \eqref{eq:x27}, \eqref{eq:x37} and \eqref{eq:x38} represent the average scale factor $a$, cosmological constant $\lambda$, pressure $p$, energy density $\rho$ and all are found to be the functions of cosmic time $t$. The metric potentials $A$ and $B$ tends to zero as $ t \rightarrow 0 $ suggesting that the space--time collapses whereas as $ t \rightarrow \infty $ they indicate a singularity. When $ t \rightarrow 1$ the metric potential $A$ and $B$ becomes constant i.e. space--time reduces to flat.\\

\begin{figure*}
\includegraphics[scale=0.45]{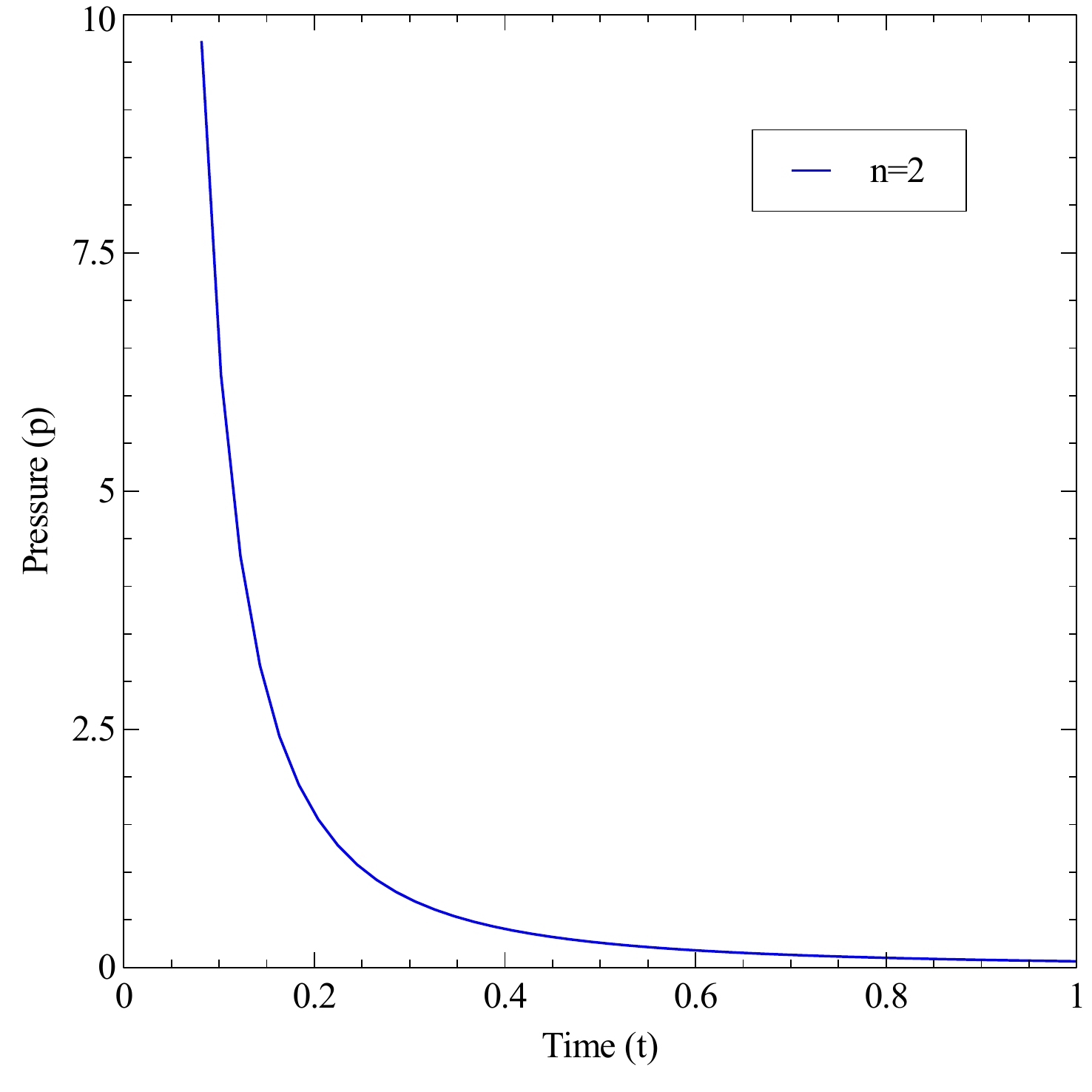}
\includegraphics[scale=0.45]{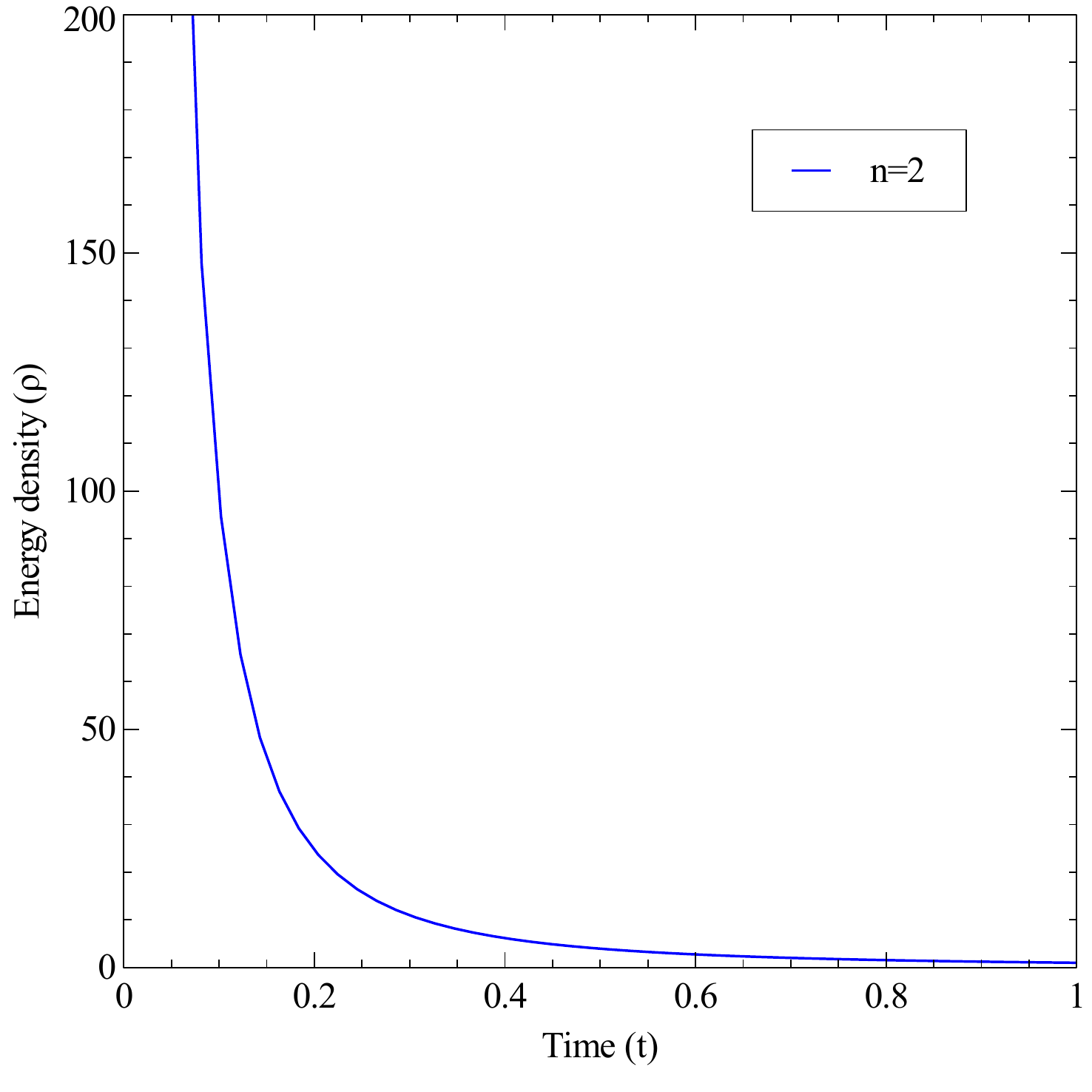}
\caption{Variations of pressue $p$ (\emph{left}) and of energy density $\rho$ (\emph{right}) as a function of time $t$. All quantities are in arbitrary units.}\label{fig3}
\end{figure*}

We observe that the directional Hubble's parameters $H_{x}=H_{y}$, $H_{z}$ as well as mean generalized Hubble's parameter $H$ are functions of cosmic time $t$ and that they tends to decrease with increase in time which is graphically shown in the left plot of figure~\ref{fig1} for arbitrary time. The right plot shows the behavior of spatial volume $V$ as a function of $t$ and reveals that $V$ increases as cosmic time tends to infinity and vanishes at $t=0$, contrary to what we observed for $H$. Figure~\ref{fig2} shows time dependent evolution of expansion scalar $\theta $ and shear scalar $\sigma$. The figure reveals that these parameters tends to zero as cosmic time tends to infinity. The energy density $\rho$ and pressure $p$ shown in fig.~\ref{fig3} also diverges as $t\rightarrow 0$ and becomes zero as $t \rightarrow \infty$.

Since $\frac{\sigma^{2}}{\theta^{2}}\neq 0$, in such a case the model does not approach isotropy. The mean anisotropy parameter $\Delta$ is constant throughout the evolution of the universe as it does not depend on the cosmic time. 
%\begin{center}

\section{Acknowledgement}
%\end{center}
PKA would like to acknowledge the Department of Science and Technology, New Dehli, India for providing INSPIRE fellowship.

\bibliographystyle{raa}\bibliography{mybibliography}
\end{document}